\documentclass[reprint,amsmath,amssymb,aps]{revtex4-1}
\usepackage{dcolumn}% Align table columns on decimal point
\usepackage{bm}% bold math
\usepackage{mathrsfs}
 \usepackage{balance} 

\usepackage[colorlinks,linkcolor=blue,anchorcolor=blue,citecolor=blue]{hyperref}

\usepackage[titletoc]{appendix}
\usepackage{graphicx}
\usepackage{tikz}

\begin{document}

\title{Gravity-induced geometric spin Hall effect as a probe  of universality of free fall of quantum particle}
\author{Zhen-Lai Wang}
\affiliation{School of  Mathematics and Physics,
Hubei Polytechnic University, Huangshi 435003, China}
\author{Xiang-Song Chen}%
\email{cxs@hust.edu.cn}
\affiliation{School of Physics and
MOE Key Laboratory of Fundamental Quantities Measurement,
Huazhong University of Science and Technology, Wuhan 430074, China}

\date{\today}
\begin{abstract}
We discuss a novel gravitational effect that the space-averaged free-fall point of quantum particle undergoes a spin-dependent transverse shift in the Earth's gravitational field. This effect is  similar to the geometric spin Hall effect (GSHE) [Aiello {\it et al.}, Phys. Rev. Lett. 103, 100401 (2009)], and can be called gravity-induced GSHE. To some extent, this effect can be viewed as a kind of violation of the universality of free fall (UFF) or weak equivalence principle (WEP) in the quantum domain. 
\end{abstract}

\maketitle

\section{Introduction}

The universality of free fall (UFF) is tested as a weak form of the Einstein equivalence principle, which is the most important guide when establishing Einstein's general relativity. The classical tests of UFF with macroscopic masses have achieved a high precision of about $10^{-13}$~\cite{Will04,Schl08}, and no violations were observed so far. To extend the domain of the test body, verifications of UFF based on microscopic particles in the quantum regime have been studied theoretically and experimentally since 1960's~\cite{Ni10}. Recently, WEP-test experiments using atom interferometers were proposed to reach the level of $10^{-15}$~\cite{Agui14,Will16}.  Quantum systems are advantageous in testing WEP with regard to  fundamental properties such as charge~\cite{Ditt04}, matter/anti-matter~\cite{ALPHA13,Hohe13,Hami14}, spin~\cite{Tara14,Duan16,Rour17} and internal structures. 
Possible violations of equivalence principle were discussed extensively, such as by spin-gravity coupling ~\cite{Mash00,Silen07,Obuk09,Obuk16}, by spin-torsion coupling ~\cite{Heh76,Shap02,Obuk14}, in extended or modified theories of gravity, and in almost all tentative theories to unify general relativity and the standard model of particle physics~\cite{Damo12,Lamm03}. 

Theoretical investigations have offered  a wide variety of approaches to the WEP in the quantum domain~\cite{Cole,Alva,Viol,Chow,Rosi17,Orla,Anas,Schw19,Flor}. However, quantum particles differ critically from classical point-like particles in many respects, because of  their wave-like features and inherent spacial extension.  Even the notion of  WEP for quantum systems is not very clear and it may be different from the conventional WEP for classical systems ~\cite{Davies, Ali, Acci, Mous,Herr12,Zych15}.  In this paper, considering the spatial extension of matter waves, we reveal an interesting phenomenon that the space-averaged free-fall point of quantum particles allows a spin-dependent transverse split in the gravitational field. Since such an effect is similar to the geometric spin Hall effect (GSHE) discussed in Refs.~\cite{Aiel09,Korg14}, we call it gravity-induced GSHE.

For a light beam, the GSHE states that a spin-dependent transverse displacement of the light intensity centroid is observed in a  plane tilted with respect to the propagation direction. Unlike the conventional spin Hall effect of light as a result of light-matter interaction~\cite{Onod04,Host08, Blio13}, the GSHE of light is of purely geometric nature. Analogously, the gravity-induced GSHE differs from the so-called gravitational Hall effect presented in the literature~\cite{Goss07,Hacy12,Yama17, Oanc20} which describes a helicity-dependent geodesic line.  

In this paper, we discuss the gravity-induced GSHE with spin-polarized electron beams. (For light beams the gravity-induced GSHE is too small in terrestrial experiments and is not considered here.) Our discussion applies equally to other matter waves such as spin-polarized neutron and atom beams. The paper is organized as follows. First, we derive an approximate wave-packet solution of the covariant Dirac equation in the Newtonian limit. With this solution, we demonstrate the gravity-induced GSHE of freely-falling electrons. Next, we present an alternative derivation of the gravity-induced GSHE in a simple method without employing detailed knowledge of the wave-function of electron beam. Then, we go further to discuss more configurations which can display gravity-induced GSHE. Finally, we  give our conclusions.

\section{ Gravity-induced GSHE}
\begin{figure}[htp]
\centering
\begin{tikzpicture}[scale=0.85]
 \draw[fill=green!20](4,-0.8) -- (8.4,-1.4) -- (8,-0.14) -- (3.62,0.7) -- cycle;
  \draw[fill=gray!20](4,-0.8)--(3.62,0.7)--(3.6,0.12)--(3.98,-1.39)--(8.4,-1.4) -- cycle;
     \draw(4,-0.8) -- (4,-1.4);
   	  \draw (6.9,-0.8) node[left,blue!80] {$K^{\prime}$};
     \draw[dashed,line width=0.4pt,-latex,blue!60] (3.85,0.0)--(8.6,-0.88)node[above,blue!80] {{\scriptsize $x^{\prime}$}};	
	 \draw[dashed,line width=0.5pt,-latex,blue!60](5.75,0.4) -- (6.16,-1.75) [right]node{{\scriptsize $y^{\prime}$}};
	%---------------------detecting plane%
	 \shadedraw[inner color=red,outer color=white,draw=white,opacity=0.8] (-0.8+7-0.4,3-0.3)  circle (11 pt);
	 \draw (6-0.4,2.8-0.3) node[left,blue!80] {$K$};
	%-----------------------particle%	
    \draw[line width=0.4pt,-latex,blue!60] (5.8,3-0.3)--(0.6+7-0.4,3-0.3)[above]node{{\scriptsize $x$}};
	\draw[line width=0.4pt,-latex,blue!60] (5.8,3-0.3)--(-1.3+7-0.4,2-0.3)[anchor=east]node{{\scriptsize $y$}};
	\draw[line width=0.4pt,-latex,blue!60] (5.8,3-0.3)--(5.8,0.9);
	\draw(5.83,1) node[anchor=east,blue!60]{{\scriptsize $z$}};
	%---------------------coordinate axis%
	\draw[-latex,line width=0.4pt,red!60] (6.1,1.8) -- (6.1,1);
	\draw	(6,1.4) node[right,red!80]{{\scriptsize $g$}};
		%---------------------gravity%
	\draw[-latex,line width=0.4pt,blue!60] (5.97,2.7-0.3) -- (5.97,2.3-0.4);
	\draw	(5.86,2.6-0.4) node[right,blue!80] {$\bm{s}$};	
	%---------------------spin%
	
	  \shadedraw[inner color=red,outer color=white,draw=red!40,opacity=0.8,rotate=-5] (5.9,0.12) ellipse [x radius=0.5cm, y radius=0.3cm];  %-------------------end of trajectory%
	\draw[-latex,line width=0.4pt,blue!60] (5.98,0.8) -- (5.98,-0.2);	
\draw	(5.93,0.5)node[right,blue!80] {$\bm{s}$};	
	\draw	(5.93,-0.05)node[right,blue!80]{\scriptsize $\bm{P_g}$};
			%---------------------momentum and spin%

   \draw[line width=0.4pt,blue!80]  (6.8,-1.2) edge[bend right=30,looseness=0.8] (6.8,-1.4);
    \draw	(6.8,-1.26) [left]node[blue!80]{{\scriptsize $\theta$}};	

\end{tikzpicture}

\caption{A schematic set-up to display gravity-induced GSHE. Static electrons carrying spin along the $z$ axis are released to fall freely and hit a detection plane $x^{\prime}$-$y^{\prime}$ tilted by an angle $\theta$ with respect to the horizontal plane. The detection and beam frames are denoted by $K^{\prime}$ and  $K$, respectively.}\label{pic1}
\end{figure}

We consider a simple system of electron originally carrying spin along the vertical direction ($z$-axis) and falling freely from rest from a height $h_g$ towards a tilted detection plane [see Fig.~\ref{pic1}]. The electron's space-averaged point of free fall will be shifted along the $y$ axis by an amount $\delta\sim({\lambda_g}/{4\pi})\sigma\tan\theta$ compared to its classical point of free fall. Here $\sigma=\pm1/2$ is the initial polarization of electron along the positive $z$-axis. This displacement $\delta$ is an order of magnitude smaller than the \textit{de Broglie} wavelength of the electron when it hitting the detector $\lambda_g\sim 2\pi\hbar/(m\sqrt{2gh_g})$. This effect implies that electrons or other quantum particles with different spin orientations follow ``different paths", and in this sense can be viewed as a kind of violations of UFF.

\subsection{Heuristic result from GSHE}

To calculate the space-averaged free-fall point of electron, what we need to know is  the spatial distribution of the electron beam's intensity in the detection plane. Following the method of Aiello {\it et al.}~\cite{Aiel09}, we choose the energy flux of the electron beam to represent its intensity and so consider the energy-momentum (E-M) tensor $T^{\mu\nu}$ of the electron beam. Thus, the space-averaged free-fall point of electrons can be calculated as the barycenter of the energy flux $T^{\prime z0}$ across the tilted detection plane: 
\begin{equation}\label{YK}
{\langle y \rangle}_{g}=\int y^{\prime}~T^{\prime z0}\,\mathrm{d}x^{\prime}\mathrm{d}y^{\prime}{\Big /}\int ~T^{\prime z0}\,\mathrm{d}x^{\prime}\mathrm{d}y^{\prime}.
\end{equation}
Note that the energy flux $T^{\prime z0}$ is defined in the detection frame and not in the beam frame.

Before going into the detailed calculation, we first explain a heuristic way of understanding the gravity-induced GSHE, by making a close connection to the ordinary GSHE, which originates from a non-zero spin projection in the detection plane and has no relevance to gravity.  To deal with the gravitational interactions, we follow Refs.~\cite{Obuk01,Sile05,Mash13} and adopt two reasonable approximations: First, the spin-precession effect probed by GP-B experiment can be safely neglected in our case. Second, the motion of the electron wave-packet's center can be approximately replaced by its classical trajectory. With these two approximations, the gravity-induced GSHE can be converted to an ordinary GSHE: Gravity just induces a kinematic configuration that the particle can hit the detector with non-zero spin projection in the detection plane, and then the ordinary GSHE occurs.  Considering the electrons as set up in Fig.~\ref{pic1}, we can directly quote the expression for ordinary GSHE as derived in Ref.~\cite{Aiel09}, and write down our result for the gravity-induced GSHE: 
\begin{equation}\label{YK0}
{\langle y \rangle}_{g}\simeq\frac{\lambda_g}{4\pi}\sigma\tan\theta.
\end{equation}
Here $\lambda_g=2\pi\hbar/p_g$ is the \textit{de Broglie} wavelength with the momentum $p_g$ when the electron arriving at the detection plane. For a non-relativistic electron and in the Newton's gravitation limit, we have $p_g\sim m\sqrt{2gh_g}$ and so ${\langle y \rangle}_{g}\sim\hbar\sigma\tan\theta/(2m\sqrt{2gh_g})$. Strictly speaking, Eq.~(\ref{YK0}) is only the leading order effect. Since this effect is pretty small, throughout this paper we omit the discussion of hight-order corrections such as from the angular spread of the beam.

\subsection{ Deriving Gravity-induced GSHE with wave-function }

To convince the reader that our heuristic argument gives the correct result, we now make an explicit calculation of  the electron's motion in the gravitational field. The dynamics of electron  in a gravitational field should be described by the covariate Dirac equation in a curved spacetime,  which is given by~\cite{Birr84}
\begin{equation}\label{de}
(i{\gamma}^{a}D_{a}-m)\psi=0.
\end{equation}
Here ${\gamma}^{a}(a=0,...,3.)$ is the flat Dirac matrix, defined by ${\gamma}^a=e_{\mu}^{~a}\gamma^{\mu}$ in the local tetrad frame. Hereafter the Latin indices will denote flat indices and the Greek indices curved indices. The tetrad field can be defined by $g_{\mu\nu}=e_{\mu}^{~a}e_{\nu}^{~b}\eta_{ab}$, and we adopt the flat metric $\eta_{ab}=\rm{diag}(+,-,-,-)$. $D_{a}$ is the covariant derivative for the spinor field:  
\[
D_{a}=e^{\mu}_{~a}D_{\mu},~~~~D_{\mu}=\partial_{\mu}-\frac{i}{2}\omega_{\mu}^{~ab}S_{ab},
\]
where $S_{ab}=i[\gamma_a,\gamma_b]/4$ and $\omega_{\mu}^{~ab}=-\omega_{\mu}^{~ba}$ is the spin connection. Throughout this paper, we choose the Dirac representation of Dirac matrices: 
\[
\gamma^{\hat{0}}=\begin{pmatrix}
I&0\\
0&-I
\end{pmatrix},~~~\gamma^{\hat{j}}=\begin{pmatrix}
0&\sigma^j\\
-\sigma^j&0
\end{pmatrix},~ j=1,2,3.
\]
Hereafter, hated specific indices denote flat indices and $\sigma^j$ is the Pauli matrices.

We consider the Earth's gravitational field near it's surface in the Newtonian limit, which is described by the metric (see, e.g.,~\cite{Hacy12,Rind,Rohr,Muno})
\begin{equation}\label{metric}
{\rm d}s^2=A^2(z){\rm d}t^2-{\rm d}\bm{r}\cdot{\rm d}\bm{r},~~A(z)=1-gz.
\end{equation}
Here we choose the gravitational acceleration $g$ along the positive direction of $z$-axis,  and the physical domain is limited by the scales $L_g=c^{2}/g\approx 0.97~\rm{ly}$.  This metric describes approximately the gravitational field  in a small region on the Earth's surface and so $|gz/c^2|=|z/L_g|\ll1$ is necessary.

With this metric, the tetrads and their inverse are 
\[\label{tetrad}
e_{\mu}^{~a}=\begin{pmatrix}
A&\\
&1\\
&&1\\
&&&1
\end{pmatrix},~~e^{\nu}_{~b}=\begin{pmatrix}
A^{-1}&\\
&1\\
&&1\\
&&&1
\end{pmatrix}.
\]
Using the zero-torsion condition~\cite{Carroll}, one can derive the non-zero components of the spin connection  
$\omega_{\mu}^{~ab}$:
\[
\omega_{0}^{~\hat{3}\hat{0}}=-\omega_{0}^{~\hat{0}\hat{3}}=A^{\prime}=-g,
\]
where the prime represents the derivative with respect to $z$. Moreover, we can  deal with  the covariant derivative:
$$ D_{\hat{j}}=e^{\mu}_{~\hat{j}}D_{\mu}=D_{j}=\partial_{j},$$
$$D_{\hat{0}}=e^{\mu}_{~\hat{0}}D_{\mu}=A^{-1}D_{0}=A^{-1}(\partial_{t}+A^{\prime}\gamma^{\hat{0}}\gamma^{\hat{3}}/2).$$
Now Eq.~(\ref{de}) can be changed into~\cite{Boul,Cris}
\begin{equation}\label{DE}
(i\gamma^{\hat{0}}A^{-1}\partial_t+i\gamma^{\hat{j}}\partial_j+\frac{i}{2}\gamma^{\hat{3}}A^{-1}A^{\prime}-m)\psi=0.
\end{equation}
It can be further rewritten in a Schr{\"o}dinger-like form
\begin{equation}\label{SE}
i\partial_t\psi=A(-i \alpha^{\hat{j}}\partial_j-i\frac{A^{\prime}}{2A}\alpha^{\hat{3}}+\beta m)\psi=H\psi,
\end{equation}
where $\alpha^{\hat{j}}=\gamma^{\hat{0}}\gamma^{\hat{j}}$ and $\beta=\gamma^{\hat{0}}$.  (The Hamiltonian $H$ also can be written in an explicitly Hermitian form $H=[(-i \alpha^{\hat{j}}\partial_j)+(-i \alpha^{\hat{j}}\partial_j)A]/2+A\beta m$.) It is convenient to introduce a function substitution:
\begin{equation}\label{FS}
\psi(t,\bm{r})=A^{-1/2}\Psi(t,\bm{r}).
\end{equation}
In terms of the new spinor $\Psi$, Eq.~(\ref{SE}) can be converted into 
\begin{equation}\label{NSE}
i\partial_t\Psi=A(-i\alpha^{\hat{j}}\partial_j+\beta m)\Psi={\cal H}\Psi=A{\cal H}_{0}\Psi.
\end{equation}

We turn now to the  calculation of  energy flux density. For the Dirac field, the familiar symmetric E-M tensor~\cite{note,Wang} is 
 \begin{equation}\label{Ts}
T_{ab}=\frac{i}{4}\overline{\psi}(\gamma_a D_b+\gamma_b D_a)\psi+h.c.
\end{equation}
where $+h.c.$ indicates the addition of the Hermitian conjugate of the foregoing terms. Recalling that $ D_{\hat{j}}=\partial_{j}$ and $D_{\hat{0}}=A^{-1}(\partial_{t}+A^{\prime}\alpha^{\hat{3}}/2)$, we can get the energy flux density 
 \begin{eqnarray}\label{GT1}
T^{j0}&=&e^{j\hat{a}}e^{0\hat{b}}T_{ab}=\frac{A^{-1}}{4i}\overline{\psi}(\gamma_{\hat{0}} D_{\hat{j}}+\gamma_{\hat{j}}D_{\hat{0}})\psi+h.c.\nonumber\\
&=&\frac{A^{-1}}{4i~}\big[\psi^{\dagger}\partial_j\psi-(\partial_j\psi^{\dagger})\psi\big]
+\frac{i}{4}A^{-2}\big[\psi^{\dagger}\alpha^{\hat{j}}\partial_t\psi\nonumber\\
&~&~~-(\partial_t\psi^{\dagger})\alpha^{\hat{j}}\psi\big]
+\frac{A^{\prime}A^{-2}}{2~}\varepsilon_{3jk}\psi^{\dagger}\Sigma^{\hat{k}}\psi,
\end{eqnarray} 
where $\Sigma^{\hat{k}}=-i\varepsilon_{ijk}\alpha^{\hat{i}}\alpha^{\hat{j}}/4$. By means of Eqs.~(\ref{FS}) and  (\ref{NSE}), we obtain 
 \begin{eqnarray}\label{GT2}
T^{j0}&=&\frac{A^{-2}}{2i}\big[\Psi^{\dagger}\partial_j\Psi-(\partial_j\Psi^{\dagger})\Psi\big]
+\frac{A^{-2}}{2~}\varepsilon_{jkl}\partial_k(\Psi^{\dagger}\Sigma^{\hat{l}}\Psi) \nonumber\\
&~&+\frac{A^{\prime}A^{-3}}{2~}\varepsilon_{3jk}\Psi^{\dagger}\Sigma^{\hat{k}}\Psi.
\end{eqnarray} 

As it is well-known, the formal solution of Eq.~(\ref{NSE}) is given by
\begin{equation}
\Psi(t,\bm{x}) = e^{-i{\cal H}t}\Psi(0,\bm{x}).
\end{equation}
 Thus, we require the initial wave function $\Psi(0,\bm{x})$. For instance, consider a gaussian packet of half width $d$~\cite{Itzy}:
\begin{equation}\label{Psi}
\Psi(0,\bm{x})=\Psi_{0}^{1,2}=\varphi(x){\cal W}^{1,2}_{0},
\end{equation}
with
\begin{equation}\label{Phi}
\varphi(x)=\frac{1}{(\pi d^2)^{3/4}}e^{\frac{-\bm{x}^2}{2d^2}}.
\end{equation}
It describes an electron of spin up/down with zero expectation value of the momentum for ${\cal W}^{1}_{0}=\{1,0,0,0\}^T$/ ${\cal W}^{2}_{0}=\{0,1,0,0\}^T$ in the non-relativistic limit ($d\gg \lambda_c=\hbar/mc$). Then Eq.~(\ref{GT2}) can be  rewritten as
\begin{eqnarray}\label{GT3}
T^{j0}&=&-iA^{-2}\Psi^{\dagger}_{0}(e^{i{\cal H}t}\partial_j e^{-i{\cal H}t}-\partial_j)\Psi_0\nonumber\\
&~&~+\frac{A^{-2}}{2~}\varepsilon_{jkl}\partial_k(\Psi^{\dagger}_{0}e^{i{\cal H}t}\Sigma^{\hat{l}}e^{-i{\cal H}t}\Psi_0) \nonumber\\
&~&~+\frac{A^{\prime}A^{-3}}{2~}\varepsilon_{3jk}\Psi^{\dagger}_{0}e^{i{\cal H}t}\Sigma^{\hat{k}}e^{-i{\cal H}t}\Psi_0.
\end{eqnarray}

Now we need to obtain the products of exponential operators in the above Equation.  Expanding the exponential operators to first order in $g$ and  using the non-relativistic limit,  we obtain finally
\begin{equation}
e^{i{\cal H}t}\partial_j e^{-i{\cal H}t}-\partial_j\simeq\begin{cases}~~
~0,~~~~&~~~~j=1,2~~\\
~igt{\cal H}_{0},&~~~~j=3~~~~,~~~~~
\end{cases}
\end{equation}

\begin{align} \label{S1}
 &(\Psi^{1,2}_{0})^{\dagger}e^{i{\cal H}t}\Sigma^{\hat{1}} e^{-i{\cal H}t}\Psi_0^{1,2}\notag\\
 &\simeq \mp\Big[\frac{\zeta \sin (2 m t)}{mt}+\frac{\sin^2 ( m t)}{m^2t^2}\Big]\frac{xzt^2}{d^4}\varphi^2 \notag\\
&~~~\mp gt\Big[\frac{\sin (2m t)}{2mt}-\frac{\sin^2 ( m t)}{2m^2t^2}\Big]\frac{xt}{d^2}\varphi^2\notag\\
&~~~-\frac{gt\sin (2m t)}{2mt}\frac{iyt}{d^2}\varphi^2 \mp gt\Big[\frac{\sin^2 (mt)}{m^4 t^4}\notag\\
&\qquad~~~-\frac{\sin (2 m t)}{m^3 t^3}+\frac{\cos (2 m t)}{m^2 t^2}\Big](d^2-z^2)\frac{xt^3}{d^6}\varphi^2 ,~
 \end{align} 
\begin{align}  \label{S2}
 &(\Psi^{1,2}_{0})^{\dagger}e^{i{\cal H}t}\Sigma^{\hat{2}} e^{-i{\cal H}t}\Psi_0^{1,2}\notag\\
 &\simeq \mp\Big[\frac{\zeta \sin (2 m t)}{mt}+\frac{\sin^2 ( m t)}{m^2t^2}\Big]\frac{yzt^2}{d^4}\varphi^2 \notag\\
&~~~\mp gt\Big[\frac{\sin (2m t)}{2mt}-\frac{\sin^2 ( m t)}{2m^2t^2}\Big]\frac{yt}{d^2}\varphi^2\notag\\
&~~~+\frac{gt\sin (2m t)}{2mt}\frac{ixt}{d^2}\varphi^2 \mp gt\Big[\frac{\sin^2 (mt)}{m^4 t^4}\notag\\
&\qquad~~~-\frac{\sin (2 m t)}{m^3 t^3}+\frac{\cos (2 m t)}{m^2 t^2}\Big](d^2-z^2)\frac{yt^3}{d^6}\varphi^2,
 \end{align} 
 \begin{align}  \label{S3}
 &(\Psi^{1,2}_{0})^{\dagger}e^{i{\cal H}t}\Sigma^{\hat{3}} e^{-i{\cal H}t}\Psi_0^{1,2}\notag\\
 &\simeq \sigma\varphi^2 \mp \Big[\frac{\zeta \sin (2 m t)}{mt}+\frac{\sin^2 ( m t)}{m^2t^2}\Big]\Big(2d^2\notag\\
&~\qquad~-x^2-y^2\Big)\frac{t^2}{d^4}\varphi^2 \pm gt\Big[\frac{\sin^2 (mt)}{m^4 t^4}-\frac{\sin (2 m t)}{m^3 t^3}\notag\\
&~\qquad\qquad~~~+\frac{\cos (2 m t)}{m^2 t^2}\Big](2d^2-x^2-y^2)\frac{zt^3}{d^6}\varphi^2.
 \end{align} 
The details of the above results are presented in  Appendix. For simplicity, we use the shorthand
$$\langle\Sigma^{\hat{l}}\rangle=(\Psi^{1,2}_{0})^{\dagger}e^{i{\cal H}t}\Sigma^{\hat{l}} e^{-i{\cal H}t}\Psi_0^{1,2}, ~~~l=1,2,3$$
and then can write 
 \begin{align}
T^{x0}&=\frac{A^{-2}}{2~}\big(\partial_y\langle\Sigma^{\hat{3}}\rangle-\partial_z\langle\Sigma^{\hat{2}}\rangle-gA^{-1}\langle\Sigma^{\hat{2}}\rangle\big),\label{Tx}\\
T^{z0}&=\frac{A^{-2}}{2~}\big(2mgt\varphi^2+\partial_x\langle\Sigma^{\hat{2}}\rangle-\partial_y\langle\Sigma^{\hat{1}}\rangle\big).\label{Tz}
 \end{align} 
In fact, in the non-relativistic limit $m\rightarrow \infty$, from Eqs.~(\ref{S1}-\ref{S3}), we have $\langle\Sigma^{\hat{l}}\rangle=0$ for $l=1,2$ and $\langle\Sigma^{\hat{l}}\rangle= \sigma\varphi^2$ for  $l=3$. These are exactly the non-relativistic and weak-field properties of spin-precession effect we mention before, and so the spin-precession effect can be ignored for simplicity in our discussion.

Returning to Eq.~(\ref{YK}), we now should notice the connection between the beam and detector frames ($K$ and $K^{\prime}$ frames). These two frames are connected by $x^{\prime a}=\Lambda^{a}_{~b}x^{b}$, with the transformation matrix
\begin{equation}
\Lambda^{a}_{~b}=\begin{pmatrix}
      1&0&0&0\\
     0&\cos\theta&0&\sin\theta \\
     0&0&1&0\\
      0&-\sin\theta&0&\cos\theta 
\end{pmatrix}.
\end{equation}
The energy flux density in $K^{\prime}$ frame is then  expressed by that in $K$ frame
\begin{align}\label{TT}
T^{\prime z0}(x^{\prime})&=\Lambda^{z}_{~a}\Lambda^{0}_{~b}T^{ab}(x)=T^{z0}(x)\cos \theta-T^{x0}(x)\sin\theta\notag\\
&=T^{z0}(\Lambda^{-1} x^{\prime})\cos \theta-T^{x0}(\Lambda^{-1} x^{\prime})\sin\theta, 
\end{align}
The full expression for $T^{\prime z0}(x^{\prime})$ is cumbersome, but its main part of interest is given by
\begin{equation}\label{T1}
T^{\prime z0}(x^{\prime})\simeq mgt\varphi^{\prime 2}\cos\theta+\frac{\sigma y^{\prime}}{d^2}\varphi^{\prime 2}\sin\theta,
\end{equation}
where 
$$\varphi^{\prime}(x^{\prime})=\frac{1}{(\pi d^2)^{3/4}}e^{\frac{-{\bm x^{\prime}}^2}{2d^2}}.$$
Finally, inserting Eq.~(\ref{T1})  into Eq.~(\ref{YK}) leads to
\begin{equation}\label{YK1}
{\langle y\rangle} _{g}=\frac{\hbar}{2p_g}\sigma\tan\theta =\frac{\lambda_g}{4\pi}\sigma\tan\theta,
\end{equation}
where $p_g=mgt$ and $\sigma=\pm1/2$ for spin-up and -down electrons. Notice that when the electron hits the detector  $h_g=gt^2/2$  and $p_g=mgt=m\sqrt{2gh_g}$ in the weak-field approximation. We therefore show the previous result of Eq.~(\ref{YK0}) and confirm the heuristic interpretation.

\subsection{ Deriving Gravity-induced GSHE without wave-function }

The {\it simple} spin-dependent result of Eq.~(\ref{YK1}) obtained by a lengthy calculation is not accidental. In fact, this result can be derived by a clever method without detailed knowledge of the electron wave-function.  From Eq.~(\ref{TT}),  Eq.~(\ref{YK}) can be reexpressed  by
\begin{align}\label{Yk}
{\langle y\rangle} _{g}&=\dfrac{\displaystyle\int y^{\prime}\big(T^{z0}(x)\cos\theta-T^{x0}(x)\sin\theta\big)\mathrm{d}x^{\prime}\mathrm{d}y^{\prime}}{\displaystyle\int \big(T^{z0}(x)\cos\theta-T^{x0}(x)\sin\theta\big)\mathrm{d}x^{\prime}\mathrm{d}y^{\prime}}\notag\\
&\simeq\dfrac{\displaystyle\int y\big(T^{z0}(x)\cos\theta-T^{x0}(x)\sin\theta\big)\mathrm{d}x\mathrm{d}y}{\displaystyle\int \big(T^{z0}(x)\cos\theta-T^{x0}(x)\sin\theta\big)\mathrm{d}x\mathrm{d}y}.
\end{align}
In the last step, expressing the area element from $K^{\prime}$ frame to $K$ frame does not change the main result. It might be more convenient to deduce the previous result again in $K$ frame.

Furthermore, the symmetric E-M tensor can be used to construct a conserved angular momentum tensor:
 \begin{equation}
M^{\lambda \mu \nu } = {x^\mu }T^{\lambda \nu } - {x^\nu }T^{\lambda \mu}.
\end{equation}
In $K$ frame, $T^{x0}$ can be ignored compared to $T^{z0}$ because the beam mainly carries energy along the propagation direction. Thus, the $T^{x0}$ term can be ignored for a small tilted angle $\theta$ in the denominator of Eq.~(\ref{Yk}),  and the remaining $T^{z0}$ term can be computed via the sum rule of energy:
 \begin{equation}\label{nk}
\int T^{z0}\,\mathrm{d}x\mathrm{d}y
= {\mathcal {K}}_{g}^{z}\simeq n\varepsilon_g,
\end{equation}
where $n$ is the electron number per unit  time across the plane $x$-$y$, namely the electron number flux. Thus, ${{\mathcal K}}_{g}^{z}$ denotes the energy per unit time across the plane $x$-$y$. Notice that the electron's energy $\varepsilon_g=\sqrt{m^2+p_g^2}\simeq m$ in the non-relativistic limit.  Due to the axial symmetry of the beam around its beam axis ($z$-axis), $T^{z0}$ should be even function of $x$ and $y$. Thus, the  integral of $y T^{z0}\cos\theta$ would  vanish and only the  integral of $y T^{x0}\sin\theta$ was left in the numerator of Eq.~(\ref{Yk}). Again, due to the axial symmetry of the beam, we have an angular momentum sum rule as follows
\begin{eqnarray}\label{na}
&~&\int -yT^{x0} \mathrm{d}x\mathrm{d}y=\int xT^{y0}\mathrm{d}x\mathrm{d}y\nonumber\\
&~&
=\frac{1}{2}\int (xT^{y0}-yT^{x0})\mathrm{d}x\mathrm{d}y=\dfrac{1}{2}N \sigma\hbar.
\end{eqnarray}
Here $N$ is the electron number per unit length along the direction of propagation and we have $n=v_gN=Np_g/\varepsilon_g$, where $v_g$ is the electron speed when hitting the detector. 

Substituting Eqs.~(\ref{nk}) and (\ref{na}) into Eq.~(\ref{Yk}), we can verify again that the barycenter of a spin-polarized beam's energy flux
\begin{equation}\label{Yksym}
{\langle y \rangle}_{g}=\dfrac{N\sigma\hbar\sin\theta}{2n\varepsilon_g\cos\theta}=\frac{\lambda_g}{4\pi}\sigma\tan\theta.
\end{equation}
We can offer a simple explanation of Eq.~(\ref{Yksym}). For Eq.~(\ref{YK}), the denominator is related to the component of electron's energy flux or momentum normal to the detection plane, i.e. the longitudinal energy flux or momentum $p_g\cos\theta$; the numerator corresponds to  the projection of electron's spin in the detection plane, i.e.  the transverse spin  angular momentum  $\sigma\hbar\sin\theta$. Hence, we get immediately the result ${\langle y \rangle}_{g} \propto(\lambda_g/2\pi)\sigma\tan\theta$. This means that the electrons  display spin-dependent pattern  in the tilted detection plane and their barycenter of the  free-fall point  changes with their spin orientations.  Additionally,  in comparison with the classical particle, the quantum particle with spin  polarization is able to fall freely in a different ``path structure".

\subsection{ Other configurations of Gravity-induced GSHE}
 \begin{figure}[h]
\centering
\begin{tikzpicture}[scale=0.6]
	\draw(-2.6,3.2)node{{\scriptsize $(a)$}};
	\draw[line width=0.4pt,-latex,blue!60] (-0.8,3)--(0.6,3)[above]node{{\scriptsize $x$}};
	\draw[line width=0.4pt,-latex,blue!60] (-0.8,3)--(-1.3,2)[anchor=east]node{{\scriptsize $y$}};
	\draw[line width=0.4pt,-latex,blue!60] (-0.8,3)--(-0.8,0.5)[anchor=east]node{{\scriptsize $z$}};
	\draw	(-1.4,2.5) node[above,blue!80] { $K$};	
	%---------------------coordinate system%
		\draw[-latex,line width=0.4pt,red!60] (-1,2.0) -- (-1,1.2);
		\draw	(-1.2,1.3) node[above,red!80]{{\scriptsize $g$}};
		%---------------------gravity%
	 \shadedraw[inner color=red,outer color=white,draw=white,opacity=0.8] (-0.8,3)  circle (11 pt);
	%----------------------particle%
	\draw[-latex,line width=0.4pt,red!60] (-0.6,3.1) -- (0.1,3.1);
	\draw	(-0.2,3.1) node[above,red!80] {$\bm{s}$};	
	%---------------------spin%
		\draw[-latex,line width=0.4pt,blue!60] (-0.6,1.6) -- (-0.6,0.6)node[right,blue!80] {\scriptsize $\bm{P_g}$};	
		\draw[-latex,line width=0.4pt,red!60] (-0.6,1.6) -- (0,1.6)node[above,red!80] {$\bm{s}$};	
	%---------------------momentum and spin%
	
     \draw[fill=green!20](-4,-1.5) -- (0.8,-1.5) -- (0.8+0.8*1.5,0) -- (0.8+0.8*1.5-4.8,0) -- cycle;
     \draw[dashed,line width=0.5pt,-latex,blue!60] (-3.8,-0.7)--(0.8+0.8*1.5,-0.7)[above]node{{\scriptsize $x^{\prime}$}};
	\draw[dashed,line width=0.5pt,-latex,blue!60](-0.2,0.2) -- (-1.8,-1.8) [anchor=east]node{{\scriptsize $y^{\prime}$}};
		\draw	(-1.7,-1.4) node[above,blue!80] { $K^{\prime}$};	
%---------------------detection plane%
\shadedraw[inner color=red,outer color=white,draw=red!50,opacity=0.8,rotate=3] (-0.9,-0.6) ellipse [x radius=0.6cm, y radius=0.4cm];
%-------------------end of trajectory%
%------------------------------------------------------left panel--------------------------------------------------%
 \draw(-3.7+6,3.2)node{{\scriptsize $(b)$}};
   \draw[line width=0.4pt,-latex,blue!60] (-2.5+6.0,3)--(6.0,3)[above]node{{\scriptsize $x$}};
	\draw[line width=0.4pt,-latex,blue!60] (-2.5+6.0,3)--(-3.3+6.0,2)[anchor=east]node{{\scriptsize $y$}};
	\draw[line width=0.4pt,-latex,blue!60] (-2.5+6.0,3)--(-2.5+6.0,0.5)[anchor=east]node{{\scriptsize $z$}};
		\draw	(2.9,2.5) node[above,blue!80] { $K$};	
	%---------------------coordinate system%
		\draw[-latex,line width=0.4pt,red!60] (-2.6+6.0,2.0) -- (-2.6+6.0,1.2);
		\draw	(-2.8+6.0,1.3) node[above,red!80]{{\scriptsize $g$}};
		%---------------------gravity%
	 \shadedraw[inner color=red,outer color=white,draw=white,opacity=0.8] (-2.5+6.0,3) circle (11 pt);
	%-----------------------particle%
	\draw[-latex,line width=0.5pt,red!60] (-2.3+6.0,3.1) -- (-1.2+6.0,3.1);
	\draw	(-1.8+6.0,3.1) node[above,red!80] {{\scriptsize $\bm{P}$}, $\bm{s}$};	
	%---------------------spin and momentum%
	\draw[dashed,red!80] (-2.5+6.0,3.0) parabola[bend at start] (-0.6+6.0,1.2);
	%---------------------trajectory curve%
		\draw[-latex,line width=0.5pt,blue!60] (-0.6+6.0,1.2) -- (-0.25+6.0,0.55)node[anchor=east,blue!80] {\scriptsize $\bm{P_g}$};	
		\draw[-latex,line width=0.5pt,red!60] (-0.6+6.0,1.2) -- (-0.1+6.0,1.2)node[anchor=south,red!80] {$\bm{s}$};	
	%-------------------end of trajectory%
	     \draw[fill=green!20](-4+6.7,-1.5) -- (0.8+6.8,-1.5) -- (0.8+0.8*1.5+6.8,0) -- (0.8+0.8*1.5-4.8+6.7,0) -- cycle;
     \draw[dashed,line width=0.5pt,-latex,blue!60] (-3.8+6.8,-0.7)--(0.8+0.8*1.5+6.8,-0.7)[above]node{{\scriptsize $x^{\prime}$}};
	\draw[dashed,line width=0.5pt,-latex,blue!60](-0.2+6.8,0.2) -- (-1.8+6.8,-1.8) [anchor=east]node{{\scriptsize $y^{\prime}$}};
		\draw	(5,-1.4) node[above,blue!80] { $K^{\prime}$};	
		\shadedraw[inner color=red,outer color=white,draw=red!50,opacity=0.8,rotate=3] (-0.9+6.7,-1.0) ellipse [x radius=0.6cm, y radius=0.4cm];
%---------------------detection plane%
\draw[-latex,line width=0.3pt,red!60] (0.5+6.0,1.6+0.5)-- (1.2+6.0,1.6+0.5);
\draw[dashed,-latex,line width=0.4pt,red!60] (0.5+6.0,1.6+0.5)-- (0.5+6.0,0.2+0.5);
\draw[-latex,line width=0.3pt,blue!80] (0.5+6.0,1.6+0.5) -- (1.2+6.0,0.2+0.5);
\draw[line width=0.3pt] (0.75+6.0,2.1) edge[bend left=20,looseness=1] (0.6+6.0,1.9);
\draw(0.9+6.0,2.15) [below]node{{\tiny $\theta_d$}};
\draw[line width=0.3pt] (0.65+6.0,1.3+0.5) edge[bend left=20,looseness=0.8] (0.5+6.0,1.25+0.5);
\draw(0.68+6.0,1.78) [below]node{{\tiny $\theta_g$}};
\draw[dashed,line width=0.3pt] (1.2+6.0,1.6+0.5)-- (1.2+6.0,0.2+0.5);
\draw[dashed,line width=0.3pt] (0.5+6.0,0.2+0.5)-- (1.2+6.0,0.2+0.5);
%---------------------angle relation%	  	
%------------------------------------------------------right panel--------------------------------------------------%
	   \end{tikzpicture}
\caption{(a): A sketch of static electrons carrying spin along the horizontal direction ($x$-axis) and falling freely to a horizontal detection plane $x^{\prime}$-$y^{\prime}$. (b):  A sketch of an electron beam initially propagating with longitunial spin-polarization in the horizontal direction ($x$-axis) and then received by a horizontal detection plane $x^{\prime}$-$y^{\prime}$. $\theta_d$ is the deflection angle induced by gravity and $\theta_g = \pi/2 -\theta_d$.  The detection and beam frames are denoted by $K^{\prime}$ and  $K$, respectively.}\label{pic2}
\end{figure}

In the previous configuration of gravity-induced GSHE, the gravitational action is mainly to change the momentum or the \textit{de Broglie} wavelength of electron. There are other possible configurations of the gravity-induced GSHE. Here we consider two illuminating cases. Figure\hyperref[pic2]{~\ref{pic2}(a)} depicts the  configuration that the electrons with horizontal spin polarization are initially stationary and then are released into free fall. Figure\hyperref[pic2]{~\ref{pic2}(b)} shows the  configuration that the electrons carry initial horizontal velocity as well as  horizontal spin polarization and then fall freely to a horizontal detection plane.  Repeating the above analysis and computation, we can get the result of the gravity-induced GSHE for Fig.\hyperref[pic2]{~\ref{pic2}(a)}
\begin{equation}\label{Ya}
{\langle y \rangle}_{g}^{a}=\dfrac{N\sigma\hbar}{2n\varepsilon_g}=\frac{\lambda_g}{4\pi}\sigma,
\end{equation}
and that for Fig.\hyperref[pic2]{~\ref{pic2}(b)}
\begin{equation}\label{Yb}
{\langle y\rangle}_{g}^{b}=\dfrac{N\sigma\hbar}{2n\varepsilon_g\cos\theta_g}=\frac{\lambda_g}{4\pi}\sigma /\cos \theta_g.
\end{equation}
Here the angle $\theta_g$ is given by $\tan\theta_g=p_x/p_z$.  In Fig.\hyperref[pic2]{~\ref{pic2}(b)}, the electron initially has longitudinal polarization, but it acquires a transverse polarization when arriving at the detection plane because of gravitational deflection. This is the reason that Eq.~(\ref{Yb}) shows an angle-dependence of $1/\cos \theta_g$. Under the Newtonian approximation,  the results in Eqs.~(\ref{Ya}) and (\ref{Yb}) can both be written as ${\langle y \rangle}_{g}\sim\hbar\sigma/{2p_z}=\hbar\sigma/(2m\sqrt{2gh_g})$. 

All  Eqs.~(\ref{Yksym}-\ref{Yb}) reflect two kinds of  the quantum violation of UFF. One kind is that the quantum particle with different spin  orientations fall freely along ``different paths".  The other kind is that the quantum particle with spin polarization can fall freely in a ``path structure" different from the classical one.

 \section{conclusion}
As mentioned before, we can find that the shift is an order of magnitude smaller than the \textit{de Broglie} wavelength $\lambda_g$. Thus, a feasible terrestrial experiment of such effect will have to be sensitive to a position distance about $0.1\lambda_g$. Additionally,  for the weak-field and  non-relativistic approximation to make sense, we need to guarantee that the free-fall time or height of electron is sufficiently small and the electron beam's width is much larger than the Compton wavelength $\lambda_c$. For instance, if the free-fall time $t\sim 1.0~{\rm s}\ll10^{7} {\rm s}$ or the free-fall height $h_g\sim 4.9~{\rm m}\ll L_g$ and the electron beam's width $d=1.0 ~{\rm mm}\gg \lambda_c$, we have $gt\sim 9.8~ {\rm m/s}\ll c$ and $\lambda_g/8\pi=\hbar/4mgt\sim 2.9~{\rm \mu m}$. Under these conditions, the spatial resolution of detector is few micrometers. Although such effect is tiny, we are confident it can be measured. Indeed, to our knowledge, the pixel detector for particles has made it possible to access a spatial resolution on the microscale~\cite{Andr, Akib}, even on the sub-microscale~\cite{Batt}. The practical detector can be sufficient to meet the requirement of the spatial resolution under the current status of the technology.

In conclusion,  we revealed a novel phenomenon of gravity-induced GSHE, namely, the ``free-fall points" of quantum particles (or matter waves) vary with their spin polarization. The measurement of this effect will be of great interest and importance, for it implies the possible observation of the violation of UFF in the quantum realm. In comparison, in the existing tests of UFF using atomic interferometers, the atoms are treated as matter waves {\it only} when analyzing their interaction with the probing light pulse, but are treated as classical point particles when considering their gravitational action. We encourage experimentalists to test the gravity-induced GSHE as a new probe of UFF of quantum particles,  so as to clarify the notion of quantum WEP.

\section*{acknowledges}

This work is supported by the National Natural Science Foundation of China (Grant Nos. 11535005 and 11275077).  Z.-L. Wang also gratefully acknowledges financial support from the Scientific Research Project of Hubei Polytechnic University (Project No. 20xjz02R).

\appendix*
\section{ First-order approximation of products of exponential operators}

For simplicity, we put $B=i{\cal H}_0 t$ and $\zeta=-gz$, then $i{\cal H}t=i{\cal H}_0 t-igz{\cal H}_0 t=B+\zeta B$. Let us consider the expansions:
 \begin{align}
&e^{i{\cal H}t}=\sum_{n=0}^{\infty}\frac{1}{n!}(B+\zeta B)^n\simeq e^B+E,\\
&e^{-i{\cal H}t}=\sum_{n=0}^{\infty}\frac{1}{n!}(-B-\zeta B)^n\simeq e^{-B}+F.
 \end{align}
However, it is rather complicated to expand the exponential function of two {\it noncommutative} operators $B$ and $\zeta B$~\cite{Casa}. Fortunately, our concern is
with the first order in $g$, and we can arrive at
 \begin{align*}
&E=\sum_{n=1}^{\infty}\sum_{k=1}^{n}\frac{1}{n!}B^{n-k}\zeta B^{k},\\
&F=\sum_{n=1}^{\infty}\sum_{k=1}^{n}\frac{(-1)^n}{n!}B^{n-k}\zeta B^{k}. 
\end{align*}
Noting $[B,\zeta]=-gt\alpha^{\hat{3}}$ and $[B^2,\alpha^{\hat{3}}]=0$ , we can derive
 \begin{align}
&E=\zeta Be^{B}-\frac{gt}{4}\{B,\alpha^{\hat{3}}\}e^{B}\notag\\
&\qquad-\frac{gt}{8}[B,\alpha^{\hat{3}}]\frac{e^{B}-e^{-B}-2Be^{B}}{B^2},\label{E}\\
&F=-\zeta Be^{-B}-\frac{gt}{4}\{B,\alpha^{\hat{3}}\}e^{-B}\notag\\
&\qquad-\frac{gt}{8}[B,\alpha^{\hat{3}}]\frac{e^{-B}-e^{B}+2Be^{-B}}{B^2}\label{F}.
 \end{align}

\subsection*{Part I}
It is clear that $e^{i{\cal H}t}\partial_j e^{-i{\cal H}t}-\partial_j=0$ for $j=1$ and $2$. We just consider the case $j=3$:
\begin{equation}
e^{i{\cal H}t}\partial_z e^{-i{\cal H}t}-\partial_z\simeq Ee^{-B}\partial_z+\partial_ze^BF.
\end{equation}
Noting $[B,\{B,\alpha^{\hat{3}}\}]=0$ and $\{B,B[B,\alpha^{\hat{3}}]\}=0$, we have
\[[B,\alpha^{\hat{3}}]e^{\pm B}=e^{\mp B}[B,\alpha^{\hat{3}}],  \quad    \{B,\alpha^{\hat{3}}\}e^{\pm B}=e^{\pm B}\{B,\alpha^{\hat{3}}\}.\]
Using these together with 
\begin{align}\label{zeta}
&e^B\zeta e^{-B}=\zeta+[B,\zeta]+\frac{1}{2!}[B,[B,\zeta]]+\cdots\notag\\
&~~~~~=\zeta-gt\alpha^{\hat{3}}-\frac{gt}{4B^2}(e^{2 B}-1-2B)[B,\alpha^{\hat{3}}], 
\end{align}
 we can obtain
 \begin{align*}
&Ee^{-B}=\zeta B-\frac{gt}{4}\{B,\alpha^{\hat{3}}\}-\frac{gt}{8}[B,\alpha^{\hat{3}}]\frac{1-2B-e^{-2B}}{B^2},\\
&e^BF=-\zeta B+\frac{gt}{4}\{B,\alpha^{\hat{3}}\}+\frac{gt}{8}[B,\alpha^{\hat{3}}]\frac{1-2B-e^{-2B}}{B^2}.
\end{align*}
Thus
\begin{equation}\label{app1}
e^{i{\cal H}t}\partial_z e^{-i{\cal H}t}-\partial_z\simeq Ee^{-B}\partial_z+\partial_ze^BF=igt{\cal H}_0.
\end{equation}
In fact, noting $[B,\partial_z]=0$ and $[\zeta B,\partial_z]=g B=igt{\cal H}_0$, we can crosscheck Eq.~(\ref{app1})  by considering the following expansion:
 \begin{align}
&~~~~ e^{i{\cal H}t}\partial_z e^{-i{\cal H}t}-\partial_z\notag\\
&=[B+\zeta B,\partial_z]+\frac{1}{2!}[B+\zeta B,[B+\zeta B,\partial_z]]+\cdots\notag\\
&\simeq [\zeta B,\partial_z]+\frac{1}{2!}[B,[\zeta B,\partial_z]]+\frac{1}{3!}[B,[B,[\zeta B,\partial_z]]]\cdots\notag\\
&=[\zeta B,\partial_z]=igt{\cal H}_0.
 \end{align}

\subsection*{Part II} 
Similarly,
\begin{equation}
e^{i{\cal H}t}\Sigma^{\hat{l}} e^{-i{\cal H}t}\simeq e^{B}\Sigma^{\hat{l}}e^{-B}+E\Sigma^{\hat{l}}e^{-B}+e^B\Sigma^{\hat{l}} F.
\end{equation}
With Eqs.~(\ref{E}) and  (\ref{F}), we have 
 \begin{align}
&E\Sigma^{\hat{l}}e^{-B}=\Big[\zeta Be^{B}-\frac{gt}{4}\{B,\alpha^{\hat{3}}\}e^{B}-\frac{gt}{8B^2}\big(1+2B\notag\\
&\qquad-e^{2B}\big)[B,\alpha^{\hat{3}}]e^{B}\Big]\Sigma^{\hat{l}} e^{-B},\\ 
&e^B\Sigma^{\hat{l}} F=-e^B\zeta e^{-B}e^B\Sigma^{\hat{l}} e^{-B}B-\frac{gt}{4}\{B,\alpha^{\hat{3}}\}e^B\Sigma^{\hat{l}}e^{-B}\notag\\
&\qquad-\frac{gt}{8}e^B\Sigma^{\hat{l}}e^{-B}\frac{e^{2B}-1-2Be^{2B}}{B^2}[B,\alpha^{\hat{3}}].
 \end{align}
A few  useful relations are presented here
 \begin{align}
&\{B,\alpha^{\hat{3}}\}=2t\partial_{z},\label{Ba}\\
& e^B\Sigma^{\hat{l}}e^{-B}=\Sigma^{\hat{l}}+\frac{e^{2 B}-1}{2 B}[B,\Sigma^{\hat{l}}],\\
&\alpha^{\hat{3}}e^{2B}=e^{-2B}\alpha^{\hat{3}}+\frac{e^{2B}-e^{-2B}}{2B}\{B,\alpha^{\hat{3}}\}.
 \end{align}
After a tedious calculation with the help of the above relations, we finally obtain 
\begin{align}\label{s1}
&e^{i{\cal H}t}\Sigma^{\hat{l}} e^{-i{\cal H}t}\simeq  \Sigma^{\hat{l}}+f_1(B) [B,\Sigma^{\hat{l}}]-f_2(B)[\alpha^{\hat{3}},\Sigma^{\hat{l}}]\notag\\
&\qquad\qquad\qquad-f_3(B) [B,\Sigma^{\hat{l}}]\alpha^{\hat{3}}-f_4(B) \alpha^{\hat{3}}[B,\Sigma^{\hat{l}}]\notag\\
&\qquad\qquad\qquad-f_5(B) \{B,\alpha^{\hat{3}}\}[B,\Sigma^{\hat{l}}],
 \end{align}
 where
\begin{equation}\label{f1}
f_1(B)=\frac{e^{2B}-1}{2 B}+\zeta e^{2 B},\qquad\qquad\qquad~~~~~
 \end{equation} 
 \begin{equation}
  f_2(B)=\frac{g t}{4B}\big(2Be^{2B}-e^{2 B}+1\big),\qquad\qquad~~
 \end{equation}
  \begin{equation}
  f_3(B)=\frac{g t}{4 B}\big(e^{2 B}-1\big),\qquad\qquad\qquad\qquad~~
 \end{equation}
 \begin{align} 
 &f_4(B)= \frac{g t}{8 B^2}\big(2Be^{2 B}-4Be^{-2 B}+2B\notag\\
&\qquad\qquad\qquad\qquad~~~~~-e^{-2B}-e^{2B}+2\big),~
 \end{align} 
\begin{align}\label{f5}
 &f_5(B)=\frac{g t}{16 B^3}\big(8B^2e^{2 B}+4Be^{-2 B}-4Be^{2 B}\notag\\
&\qquad\qquad\qquad~~~-4 B+e^{-2 B}+3 e^{2 B}-4\big).
 \end{align} 
These operators seem highly complicated, but we are indeed concerned with the product of inserting them between $\Psi^{\dagger}_{0} $ and $\Psi_0$.  We can check easily the following equations:
\begin{equation}\label{ps1}
\Psi^{\dagger}_{0}[B,\Sigma^{\hat{l}}]\Psi_0=\Psi^{\dagger}_{0}[\alpha^{\hat{3}},\Sigma^{\hat{l}}]\Psi_0=0,
 \end{equation}
 
\begin{equation}\label{ps2}
(\Psi^{1,2}_{0})^{\dagger}B[B,\Sigma^{\hat{l}}]\Psi_{0}^{1,2}=\begin{cases}
~~\mp t^2\varphi\partial_{x}\partial_{z}\varphi,&l=1\\
~~\mp t^2\varphi\partial_{y}\partial_{z}\varphi,&l=2\\
\pm t^2\varphi(\partial_{x}^{2}+\partial_{y}^{2})\varphi,&l=3
\end{cases}
 \end{equation}
\begin{equation}(\Psi^{1,2}_{0})^{\dagger}B[\alpha^{\hat{3}},\Sigma^{\hat{l}}]\Psi_{0}^{1,2}=\begin{cases}
t\varphi(\mp\partial_{x}+i\partial_{y})\varphi,&l=1\\
t\varphi(-i\partial_{x}\mp \partial_{y})\varphi,&l=2\\
\qquad~0,&l=3
\end{cases}
 \end{equation} 
\begin{equation}\left.\begin{gathered}\Psi^{\dagger}_{0}\alpha^{\hat{3}}[B,\Sigma^{\hat{l}}]\Psi_0\\
\parallel\\
\Psi^{\dagger}_{0}[B,\Sigma^{\hat{l}}]\alpha^{\hat{3}}\Psi_0
\end{gathered}\right>=\begin{cases}
-it\varphi\partial_{y}\varphi,&\qquad l=1\\
~it\varphi\partial_{x}\varphi,&\qquad l=2\\
\qquad~0,~&\qquad l=3
\end{cases}
 \end{equation}
\begin{equation}
\left.\begin{gathered}\Psi^{\dagger}_{0}B\alpha^{\hat{3}}[B,\Sigma^{\hat{l}}]\Psi_0\\
\parallel\\
\Psi^{\dagger}_{0}B[B,\Sigma^{\hat{l}}]\alpha^{\hat{3}}\Psi_0
\end{gathered}\right>=\begin{cases}
~mt^2\varphi\partial_{y}\varphi,&l=1\\
-mt^2\varphi\partial_{x}\varphi,&l=2\\
\qquad~0,~&l=3
\end{cases}
 \end{equation}
 \begin{equation}\label{ps6}
(\Psi^{1,2}_{0})^{\dagger}\Sigma^{\hat{l}}\Psi_{0}^{1,2}=\begin{cases}
\qquad~~0,&l=1,2\\
\pm\frac{1}{2}\varphi^2=\sigma\varphi^2,&l=3
\end{cases}
 \end{equation} 
Hence, we have
\begin{align}
&\Psi^{\dagger}_{0} e^{i{\cal H}t}\Sigma^{\hat{l}} e^{-i{\cal H}t}\Psi_0\simeq \Psi^{\dagger}_{0} \Sigma^{\hat{l}}\Psi_{0}+\Psi^{\dagger}_{0}G_1(B)B[B,\Sigma^{\hat{l}}]\Psi_0\notag\\
&\qquad-\Psi^{\dagger}_{0}G_2(B)B[\alpha^{\hat{3}},\Sigma^{\hat{l}}]\Psi_0-\Psi^{\dagger}_{0}G_3(B) \alpha^{\hat{3}}[B,\Sigma^{\hat{l}}]\Psi_0\notag\\
&\qquad-\Psi^{\dagger}_{0}G_4(B) \{B,\alpha^{\hat{3}}\}B[B,\Sigma^{\hat{l}}]\Psi_0,
 \end{align}
where
\begin{equation}\label{G1}
  G_1(B)=\frac{f_1(B)-f_1(-B)}{2 B},
 \end{equation} 
 \begin{equation}
 G_2(B)=\frac{f_2(B)-f_2(-B)}{2 B},
 \end{equation}
  \begin{equation}
  G_3(B)=f_3(B)+f_4(B),~~
 \end{equation}
\begin{equation} \label{G4}
G_4(B)=\frac{f_5(B)-f_5(-B)}{2 B}.
 \end{equation} 
 Additionally, using the non-relativistic limit $d\gg \hbar/mc$, we get
\begin{align}
& \Psi^{\dagger}_{0}B^2\Psi_0=\Psi^{\dagger}_{0}(\partial^{2}_{x}+\partial^{2}_{y}+\partial^{2}_{z}-m^2)t^2 \Psi_0\notag\\
&\qquad\qquad\simeq -m^2t^2 \varphi^2=(imt)^2 \varphi^2.
 \end{align}
Applying the correspondence $B\to imt$ to the operator functions (\ref{G1}-\ref{G4}), we therefore have
\begin{align}\label{sigma}
&\Psi^{\dagger}_{0} e^{i{\cal H}t}\Sigma^{\hat{l}} e^{-i{\cal H}t}\Psi_0\simeq \Psi^{\dagger}_{0} \Sigma^{\hat{l}}\Psi_{0}+G_1(imt)\Psi^{\dagger}_{0}B[B,\Sigma^{\hat{l}}]\Psi_0\notag\\
&~~~-G_2(imt)\Psi^{\dagger}_{0}B[\alpha^{\hat{3}},\Sigma^{\hat{l}}]\Psi_0-G_3(imt)\Psi^{\dagger}_{0} \alpha^{\hat{3}}[B,\Sigma^{\hat{l}}]\Psi_0\notag\\
&~~~-G_4(imt) \Psi^{\dagger}_{0}\{B,\alpha^{\hat{3}}\}B[B,\Sigma^{\hat{l}}]\Psi_0.
 \end{align}
Substituting  Eqs.~(\ref{Ba}) and (\ref{ps2}-\ref{ps6}) together with  Eqs.~(\ref{f1}-\ref{f5}) and (\ref{G1}-\ref{G4}) into Eq.~(\ref{sigma}), we can verify Eqs.~(\ref{S1}-\ref{S3}).


\begin{thebibliography}{99}  

\bibitem{Will04} J.G. Williams, S.G. Turyshev, and D.H. Boggs, Phys. Rev. Lett. {\bf 93}, 261101 (2004).

\bibitem{Schl08} S. Schlamminger, K.-Y. Choi, T.A. Wagner, J.H. Gundlach, and E.G. Adelberger, Phys. Rev. Lett. {\bf 100}, 041101 (2008).

\bibitem{Ni10}  For a review see, for example, W.-T. Ni, Rep. Prog. Phys. {\bf 73}, 056901 (2010).

\bibitem{Agui14} D.N. Aguilera {\it et al.}  Class. Quantum Grav. {\bf 31}, 115010 (2014).

\bibitem{Will16} J. Williams, S. Chiow, N. Yu, and H. M\"{u}ller, New J. Phys. {\bf 18}, 025018 (2016).

\bibitem{Ditt04} H. Dittus, C. L\"{a}mmerzahl, and H. Selig, ‎Gen. Rel. Gravit. {\bf 36}, 571 (2004).

\bibitem{ALPHA13}  The ALPHA Collaboration and A.E. Charman,  Nat. Commun. {\bf 4}, 1785 (2013).

\bibitem{Hohe13} M.A. Hohensee, H. M\"{u}ller, and R.B. Wiringa, Phys. Rev. Lett. {\bf 111}, 151102 (2013).

\bibitem{Hami14} P. Hamilton, A. Zhmoginov, F. Robicheaux, J. Fajans, J.S. Wurtele, and H. M\"{u}ller, Phys. Rev. Lett., {\bf 112}, 121102 (2014).

\bibitem{Tara14} M.G. Tarallo, T. Mazzoni, N. Poli, D.V. Sutyrin, X. Zhang, and G.M. Tino, Phys. Rev. Lett. {\bf 113}, 023005 (2014).

\bibitem{Duan16} X.-C. Duan, X.-B. Deng, M.-K. Zhou, K. Zhang, W.-J. Xu, F. Xiong, Y.-Y. Xu, C.-G. Shao, J. Luo, and Z.-K. Hu,  Phys. Rev. Lett. {\bf 117}, 023001 (2016).

\bibitem{Rour17} A. Roura, Phys. Rev. Lett. {\bf 118}, 160401 (2017).

\bibitem{Mash00} B. Mashhoon, Class. Quantum Grav. {\bf 17}, 2399 (2000).

\bibitem{Silen07} A.J. Silenko and O.V. Teryaev, Phys. Rev. D {\bf 76}, 061101 (2007).

\bibitem{Obuk09} Y.N. Obukhov, A.J. Silenko, and O.V. Teryaev, Phys. Rev. D {\bf 80}, 064044 (2009).

\bibitem{Obuk16} Y.N. Obukhov, A.J. Silenko, and O.V. Teryaev, Int. J. Mod. Phys. Conf. Ser. {\bf 40}, 1660081 (2016).

\bibitem{Heh76} F.W. Hehl,  Gen. Relat. Grav. {\bf 4}, 333 (1973); {\it ibid.} {\bf  5}, 491(1974); F.W. Hehl, P. von der Heyde, G.D. Kerlick, and J.M. Nester, Rev. Mod. Phys. {\bf 48}, 393 (1976).

\bibitem{Shap02} I.L. Shapiro, Phys. Rep. {\bf 357}, 113 (2002).

\bibitem{Obuk14} Y.N. Obukhov, A.J. Silenko, and O.V. Teryaev, Phys. Rev. D {\bf 90}, 124068 (2014).

\bibitem{Damo12} T. Damour, Class. Quantum Grav.  {\bf 29}, 184001 (2012).

\bibitem{Lamm03} C. L\"{a}mmerzahl, {\it Quantum Gravity} (Springer, Berlin, Heidelberg, 2003), pp. 367–394.

\bibitem{Cole} R. Colella, A. Overhauser, and S. Werner, Phys. Rev. Lett. {\bf 34}, 1472 (1975).

\bibitem{Alva} C. Alvarez and R.B. Mann, Gen. Relativ. Gravit. {\bf  29}, 245 (1997).

\bibitem{Viol} L. Viola and R. Onofrio, Phys. Rev. D  {\bf 55}, 455 (1997).

\bibitem{Chow}P. Chowdhury, D. Home, A.S. Majumdar, S.V. Mousavi, M.R. Mozaffari, and S. Sinha, Class. Quantum Grav.  {\bf 29}, 025010 (2012).

\bibitem{Rosi17} G. Rosi, G. D'Amico, L. Cacciapuoti, F. Sorrentino, M. Prevedelli, M. Zych, \v{C}. Brukner, and G.M. Tino, Nat. Commun. {\bf 8}, 15529 (2017).

\bibitem{Orla}P.J. Orlando, R.B. Mann, K. Modi, and F.A. Pollock, Class. Quantum Grav. {\bf  33}, 19LT01 (2016).

\bibitem{Anas}C. Anastopoulos and B.-L. Hu, Class. Quantum Grav.  {\bf 35}, 035011 (2018).

\bibitem{Schw19} P.K. Schwartz and D. Giulini, Class. Quantum Grav. {\bf 36}, 095016 (2019).

\bibitem{Flor} P.C.M. Flores and E.A. Galapon, Phys. Rev. A {\bf 99}, 042113 (2019).

\bibitem{Davies} P.C.W. Davies and J. Fang, Proc. R. Soc. Lond. A {\bf 381}, 469 (1982).

\bibitem{Ali} M.M. Ali, A.S. Majumdar, D. Home, and A.K. Pan, Class. Quantum Grav. {\bf 23}, 6493 (2006).

\bibitem{Acci} A. Accioly and R. Paszko,  Phys. Rev. D {\bf 78}, 064002 (2008).

\bibitem{Herr12} S. Herrmann, H. Dittus, C. L\"{a}mmerzahl, Class. Quantum Grav. {\bf 29}, 184003 (2012).

\bibitem{Mous} S.V. Mousavi, A.S. Majumdar, and D. Home, Class. Quantum Grav. {\bf 32}, 215014 (2015).

\bibitem{Zych15} M. Zych and \v{C}. Brukner, Nature Phys. {\bf 14}, 1027 (2018); arXiv:1502.00971 (2015); M. Zych, {\it Quantum Systems under Gravitational Time Dilation} (Springer, Cham, 2017)


\bibitem{Aiel09} A. Aiello, N. Lindlein, C. Marquardt, and G. Leuchs, Phys. Rev. Lett. {\bf 103}, 100401 (2009).

\bibitem{Korg14} J. Korger, A. Aiello, V. Chille, P. Banzer, C. Wittmann, N. Lindlein, C. Marquardt, and G. Leuchs, Phys. Rev. Lett. {\bf 112}, 113902 (2014).


\bibitem{Onod04} M. Onoda, S. Murakami, and N. Nagaosa, Phys. Rev. Lett. {\bf 93}, 083901 (2004).

\bibitem{Host08} O. Hosten and P. Kwiat, Science {\bf 319}, 787 (2008).

\bibitem{Blio13} K.Y. Bliokh and A. Aiello, J. Opt. {\bf 15}, 014001 (2013).

\bibitem{Goss07} P. Gosselin, A. B\'{e}rard, and H. Mohrbach, Phys. Rev. D {\bf 75}, 084035 (2007).

\bibitem{Hacy12} S. Hacyan, Phys. Rev. D {\bf 85}, 024035 (2012).

\bibitem{Yama17} N. Yamamoto,   arXiv:1708.03113 (2017).

\bibitem{Oanc20}M.A. Oancea, J. Joudioux, I.Y. Dodin, D.E. Ruiz, C.F. Paganini, and L. Andersson, Phys. Rev. D {\bf 102}, 024075 (2020).


\bibitem{Obuk01} Y.N. Obukhov, Phys. Rev. Lett. {\bf 86}, 192 (2001).

\bibitem{Sile05} A.J. Silenko and O.V. Teryaev,  Phys. Rev. D {\bf 71}, 064016 (2005).

\bibitem{Mash13} B. Mashhoon and Y.N. Obukhov, Phys. Rev. D {\bf 88}, 064037 (2013).

\bibitem{Birr84}  N.D. Birrell and P.C.W. Davies, {\it Quantum Fields in Curved Space} (Cambridge University Press, Cambridge, 1984).

\bibitem{Rind} W. Rindler, {\it Relativity: Special, General, and Cosmological}, 2nd ed.(Oxford University Press, Oxford, 2001)

\bibitem{Rohr} F. Rohrlich, Ann. Phys. (N.Y.) {\bf 22}, 169 (1963).

\bibitem{Muno} G. Mu\~{n}oz and P. Jones, Am. J. Phys. {\bf 78}, 377 (2010).

\bibitem{Carroll} S.M. Carroll, {\it Spacetime and Geometry: An Introduction to General Relativity} (Addison Wesley, San Francisco, 2004)

\bibitem{Boul}N. Boulanger, P. Spindel, and F. Buisseret, Phys. Rev. D {\bf 74}, 125014 (2006).

\bibitem{Cris} L.C.B. Crispino, A. Higuchi, and G.E.A. Matsas, Rev. Mod. Phys. {\bf 80}, 787 (2008).

\bibitem{Dari}C. Dariescu and M.-A. Dariescu, Chin. Phys. Lett. {\bf 33}, 020201 (2016).

\bibitem{note} It should be noted that the E-M tensors have various versions, e.g. the canonical one and symmetric one. It may be tricky to pick out a proper one in actual application. Fortunately, the use of different E-M tensors does not change qualitatively the key features of the GSHE.

\bibitem{Wang}Z.-L Wang  and  X.-S. Chen, Phys. Rev. A {\bf 99}, 063832 (2019).

\bibitem{Itzy} C. Itzykson and J.-B. Zuber. {\it Quantum field theory }(McGraw-Hill, New York, 1980); H.C. Ohanian,  Am. J. Phys. {\bf  54}, 500 (1986).

\bibitem{Casa}F. Casas, A. Murua, and M. Nadinic, Comput. Phys. Commun. {\bf 183}, 2386 (2012).

\bibitem{Andr}L. Andricek, J. Caride, Z. Dolezal, Z. Drasal, S. Esch {\it et al.}, Nucl. Instrum. Meth. A {\bf 638}, 24 (2011).
\bibitem{Akib}K. Akiba, M. Artuso, R. Badman, A. Borgia, R. Bates {\it et al.}, Nucl. Instrum. Meth. A {\bf 661}, 31 (2012).
\bibitem{Batt}M. Battaglia, D. Bisello, D. Contarato, P. Denes, P. Giubilato, S. Mattiazzo, D. Pantano, and S. Zalusky, Nucl. Instrum. Meth. A {\bf 654}, 258 (2011).

\end{thebibliography}
\end{document}